\documentclass[fleqn,usenatbib]{mnras}

\usepackage{newtxtext,newtxmath}

\usepackage[T1]{fontenc}

\DeclareRobustCommand{\VAN}[3]{#2}
\let\VANthebibliography\thebibliography
\def\thebibliography{\DeclareRobustCommand{\VAN}[3]{##3}\VANthebibliography}

\usepackage{graphicx}	
\usepackage{amsmath}	

\title[Z Cam shell expansion and age]{Introducing the Condor Array Telescope: III. The expansion and age of the shell of the dwarf nova Z Camelopardalis, and detection of a second, larger shell}

\author[M. M. Shara et al.]{
Michael M. Shara,$^{1}$\thanks{E-mail: mshara@amnh.org (MMS)}
Kenneth M. Lanzetta,$^{2}$
James T. Garland,$^{1}$
Stefan Gromoll,$^{3}$
David Valls-Gabaud,$^{4}$
\newauthor 
Frederick M. Walter,$^{2}$
John F. Webb,$^{5}$
David R. Zurek,$^{1}$
Noah Brosch,$^{6}$
and R. Michael Rich$^{7}$
\\
$^{1}$Department of Astrophysics, American Museum of Natural History, CPW \& 79th street, New York NY 10024-5192, USA\\
$^{2}$Department of Physics and Astronomy, Stony Brook University, Stony Brook, NY 11794-3800, USA\\
$^{3}$Amazon Web Services, 410 Terry Ave. N, Seattle, WA 98109, USA\\
$^{4}$Observatoire de Paris, LERMA, CNRS UMR 8112, 61 Avenue de l'Observatoire, 75014 Paris, France\\
$^{5}$Institute of Astronomy, University of Cambridge, Madingley Road, Cambridge CB3 0HA, United Kingdom\\
$^{6}$The Wise Observatory and the Raymond and Beverly Sackler School of Physics and Astronomy, The Faculty of Exact Sciences,
Tel Aviv University, Tel Aviv 69978, Israel\\
$^{7}$Division of Astronomy, Department of Physics and Astronomy, UCLA, 430 Portola Plaza, Box 951547, Los Angeles, CA 90095-1547, USA\\
}

\date{Accepted XXX. Received YYY; in original form ZZZ}

\pubyear{2023}

\begin{document}
\label{firstpage}
\pagerange{\pageref{firstpage}--\pageref{lastpage}}
\maketitle

\begin{abstract}
The existence of a vast nova shell surrounding the prototypical dwarf nova Z Camelopardalis (Z Cam) proves that some old novae undergo metamorphosis to appear as dwarf novae thousands of years after a nova eruption. The expansion rates of ancient nova shells offer a way to constrain both the time between nova eruptions and the time for post-nova mass transfer rates to decrease significantly, simultaneously testing nova thermonuclear runaway models and hibernation theory. Previous limits on the expansion rate of part of the Z Cam shell constrain the inter-eruption time between Z Cam nova events to be $>$ 1300 years. Deeper narrow-band imaging of the ejecta of Z Cam with the Condor Array Telescope now reveals very low surface brightness areas of the remainder of the shell. A second, even fainter shell is also detected, concentric with and nearly three times the size of the ``inner'' shell. This is the first observational support of the prediction that concentric shells must surround the frequently-erupting novae of relatively massive white dwarfs. The Condor images extend our Z Cam imaging baseline to 15 years, yielding the inner shell's expansion rate as $v = 83 \pm 37$\,km\,s$^{-1}$ at 23 degrees South of West, in excellent agreement with our 2012 prediction. This velocity corresponds to an approximate age of $t = 2672^{-817}_{+2102}$ yr. While consistent with the suggestion that the most recent nova eruption of Z Cam was the transient recorded by Chinese Imperial astrologers in the year 77 BCE, the age uncertainty is still too large to support or disprove a connection with Z Cam.

\end{abstract}




\section{Introduction}
Cataclysmic variables (CVs) all comprise a white dwarf (WD) accreting matter from its binary companion \citep{Warner1995}. While the donor stars can be helium WDs, red giants or subgiants, the vast majority of known CVs contain approximately solar composition main sequence star or brown dwarf secondaries. The recent accretion rate $dM/dt$ determines the subtype of CV that we observe, while the history of $dM/dt$ over multiple Gyr drives the binary's long-term evolution. 

The continued accretion of hydrogen-rich matter onto a WD results in an increasingly dense and massive envelope supported by degenerate electron pressure. Nuclear reactions and their concomitant heating near the accreted envelope base do not, at first, lead to envelope expansion because degenerate electron pressure is temperature independent. Once the nuclear heating timescale becomes shorter than the timescale required to carry away heat \citep{Shara1981} a thermonuclear runaway ensues. The resulting classical nova \citep{Starrfield1972,Prialnik1978} fuses several percent of its envelope's hydrogen to helium in a few minutes, generating enough energy to eject the accreted envelope and achieving peak luminosities up to $10^{6} L_\odot$. Nova eruptions cease when all or most of the accreted envelope has been ejected \citep{Prialnik1979}.

Nova shells are expected to remain bright enough to be detected for multiple centuries \citep{Tappert2020} after eruption. The likely six-century-old shell of Nova Sco 1437 CE \citep{Shara2017b} is easily imaged with 1-m-class telescopes. A recent H$\,\alpha$ imaging survey of 47 CVs focused on (high mass transfer rate) nova-like variables, but did not detect any new nova shells \citep{Sahman2022}. We will return to this important negative result in Section 3, where we demonstrate that significantly lower surface brightness imaging than has hitherto been accomplished is essential to detecting the oldest nova ejecta. 

The envelope masses required to initiate a nova thermonuclear runaway on a WD range from $\sim 10^{-7} M_\odot$ for WDs close to $1.4 M_\odot$, to as large as $\sim 10^{-3} M_\odot$ for CV WDs in the $0.6 M_\odot$ range \citep{Yaron2005}. The time required to accrete a critical envelope mass (i.e.\ the time between successive nova eruptions) on $\sim 1.4 M_\odot$ WDs can be as short as 1 yr \citep{Darnley2014}, to longer than 1 Myr on low mass WDs \citep{Yaron2005}. Because the critical envelope masses are all much smaller than the masses of CV donors, all CVs, including those never having been observed to undergo a nova eruption, will eventually undergo (many) nova eruptions (see also \citealt{Ford1978}). Every CV not currently undergoing a nova eruption is detached \citep{Hillman2020} or re-accumulating an envelope onto its WD in preparation for the next nova event. 

Large $dM/dt$ (typically $> 10^{-8} M_\odot$ yr$^{-1}$) are seen in CVs with orbital periods $P \sim 5-10$ hr; these are the so-called nova-like binaries. In the century before an eruption, and for decades afterwards, most novae are indistinguishable from nova-like binaries, exhibiting high $dM/dt$ \citep{Collazzi2009}. The accretion disks of CVs exhibiting $dM/dt$ slightly less than those of nova-like variables are unstable, and undergo transitions from a hot to a cold state \citep{Lasota2001,Dubus2018}. This leads to the rapid, episodic dumping of much of the accretion disk onto the WD---a dwarf nova (DN) outburst \citep{Osaki1974}. The liberation of gravitational potential energy by the sudden accretion of most of the mass in an accretion disk ($\sim 10^{-10} M_\odot$) can liberate a few $L_\odot$ for about a week. 

The subclass of CVs with $dM/dt$ close to the nova-like-DN boundary are named after the prototype Z Camelopardalis (Z Cam). When Z Cam's $dM/dt$ from the secondary star into the accretion disk surrounding the WD is too large to produce dwarf nova outbursts, the system resembles a nova-like variable. Z Cam returns to undergoing DN outbursts when $dM/dt$ declines below the critical level \citep{Osaki1974,Buat-Menard2001,Simonsen2014}.

The hibernation scenario of CVs \citep{Shara1986,Prialnik1986,Livio1987,Kovetz1988,Hillman2020}
predicts that the mass lost during a nova eruption should lead to an increased separation of the WD from its donor, and a concomitant decrease in $dM/dt$ several centuries after an eruption, once irradiation from the post-nova, cooling WD fades. If this prediction is correct then some old novae, initially observed to be nova-like variables, should begin to undergo dwarf nova eruptions as $dM/dt$ declines. The best-known example is GK Persei (nova Per 1901) which began DN outbursts 65 years after its nova outburst. An even more extreme case is V1017 Cen, which underwent a DN event that may have triggered WD surface nuclear burning just 14 years after its classical nova outburst in 2005 \citep{Aydi2022}. \citet{Kato2022} lists 10 old novae that are currently undergoing DN eruptions.

The hibernation scenario also predicts that some DN should still be surrounded by the ejecta of their last nova eruption. The nova shell of the prototypical dwarf nova Z Cam was the first discovered \citep{Shara2007}, and it remains the largest and oldest such shell known \citep{Shara2012}. As of 2023, three dwarf novae are known to be surrounded by the ejecta of classical novae \citep{Miszalski2016,Shara2017a,Shara2017b}. 

A comparison of narrow-band H$\,\alpha+[\ion{N}{II}]$ images of the Z Cam shell, taken with the Mayall 4-m telecope in 2007 and 2010, constrained but did not detect expansion of the nova shell \citep{Shara2012}, with an upper limit of 0.17 arcsec yr$^{-1}$. Including the important effect of deceleration as the ejecta sweeps up interstellar matter in its snowplow phase, the observed upper limit on shell expansion constrains the nova event to have occurred at least 1300 years ago. This was the first strong test of the prediction of nova thermonuclear runaway theory that the inter-outburst times of most classical novae are longer than 1000 years. 

At its Gaia-determined distance of just 214 pc \citep{Bailer-Jones2021}, Z Cam must achieve a peak visual magnitude in the range $\sim +1$ to $-3$ when it undergoes a nova eruption, and it is brighter than 2nd magnitude for 2 to 3 weeks. The intriguing suggestions that Z Cam was a bright nova recorded by Chinese imperial astrologers in October and November 77 BCE \citep{Johansson2007} or more recent transients in Ursa Major \citep{Warner2015} is consistent with our expansion upper limit (see \citet{Hoffmann2019} for a critique of the latter suggestion). We predicted \citep{Shara2012} that if Z Cam is indeed the nova of 77 BCE, its ejecta should currently be expanding at $\sim 85$ km s$^{-1}$, corresponding to 110 mas yr$^{-1}$. Since 15 years have passed since our 2007 epoch of observations, and a shell expansion of 15 yr $\times$ 110 mas yr$^{-1}$ = 1.65 arcsec should be straightforward to measure, we undertook a program of narrow-band imaging of Z Cam as part of the science verification of the new Condor Array Telescope \citep{Lanzetta2023}, hereafter Paper I.

In Section 2 we briefly describe the Condor telescope, emphasising its features optimised for extremely low surface brightness imaging. We also describe the two epochs' images that we used to search for shell expansion. In Section 3 we present the Condor images to demonstrate the telescope's remarkable low surface brightness sensitivity. The methodology used to compare the two epochs' images is described in Section 4, and the measured angular and spatial expansion velocities are presented. Our results are briefly summarized in Section 5.

\section{Observations}

Because this is one of the first papers to present scientific results based on the Condor Array Telescope, we briefly describe Condor and its capabilities before describing the datasets used in our analyses.
\subsection{The Condor Array Telescope}
\label{sec:condor} 
Condor is an ``array telescope'' that consists of six apochromatic refracting telescopes of objective diameter 180 mm, each equipped with a large-format (9576 × 6388 pixels), very low read-noise (1.2 e$^{-1}$), very rapid read-time (< 1 s) CMOS camera. It is of particular relevance here that Condor is equipped with a set of $\ion{He}{II}$ 468.6, $[\ion{O}{III}]$ 500.7, $\ion{He}{I}$ 587.5, H$\,\alpha$ 656.3, $[\ion{N}{II}]$ 658.4, and $[\ion{S}{II}]$ 671.6 narrow-band filters (one per telescope). Condor has no interferometric capability; its six co-aligned refractors enable the acquisition of images of a single field through up to six different filters simultaneously. Condor is located at the Dark Skies New Mexico observatory new Animas, New Mexico. Details of the telescope are described in \citet{Lanzetta2023}.

\subsection{The Data}
\label{sec:data} 
We observed Z Cam using the Mosaic CCD camera on the Kitt Peak National Observatory (KPNO) Mayall 4-m telescope in 2007 and using Condor in 2021. 

The Kitt Peak National Observatory (KPNO) Mayall 4-m telescope with the Mosaic CCD camera at prime focus was used to observe Z Cam on the nights of 2007 January 23 and 24. The Mosaic CCD camera incorporates eight 2048 × 4096 pixel$^2$ SITe thinned CCDs, which were used with 2 × 2 on chip binning to yield an effective 8192 × 8192 pixel$^2$ format. At prime focus, the camera obtains an image scale of 0.26 arcsec per binned pixel, which results in a 36 × 36 arcmin$^2$ field of view. We obtained 106 images of total exposure time 45,300 s through the KPNO k1009 narrow-band filter. In the f/3.1 beam of the Mayall telescope the filter's wavelength centroid is 656.3 nm and its FWHM 8.1 nm, which is sensitive to the sum of H$\,\alpha$ 656.3 nm and $[\ion{N}{II}]$ 658.4 nm. The telescope was dithered between exposures.

The images were bias subtracted and flat fielded using standard procedures, and the individual images were aligned using \textsc{daophot} \citep{Stetson1987}. The images were then stitched together using \textsc{montage2}, which is a mosaicking program within \textsc{daophot}. See \citep{Shara2012} for further details, and the images of Z Cam. 

We used Condor with its complement of narrow-band filters to observe Z Cam in November 2021. The Condor CMOS cameras obtain an image scale of 0.85 arcsec per pixel, which results in the same 2.3 × 1.5 deg$^2$ (134 × 91 arcmin$^2$) field of view for each of the six telescopes. Each exposure was 600 s in length, and we obtained a total exposure time of 63,600 s through each filter with one of the six telescopes. The array was dithered by a random offset of 15 arcmin between exposures. Images of the dusk and dawn twilight sky were obtained every night, and bias observations were obtained at the end of every night. 

The observations were processed through the Condor data pipeline (Papers I and II), with steps involving bias subtraction, flat fielding and background subtraction, astrometric calibration, and photometric calibration. It is of particular relevance here that the astrometric calibrations exhibit systematic differences between the transformed pixel and celestial coordinates of $ < 0.1$ arcsec.

\section{Extended nebulosity}

\begin{figure*}
    \includegraphics[width=0.8\textwidth]{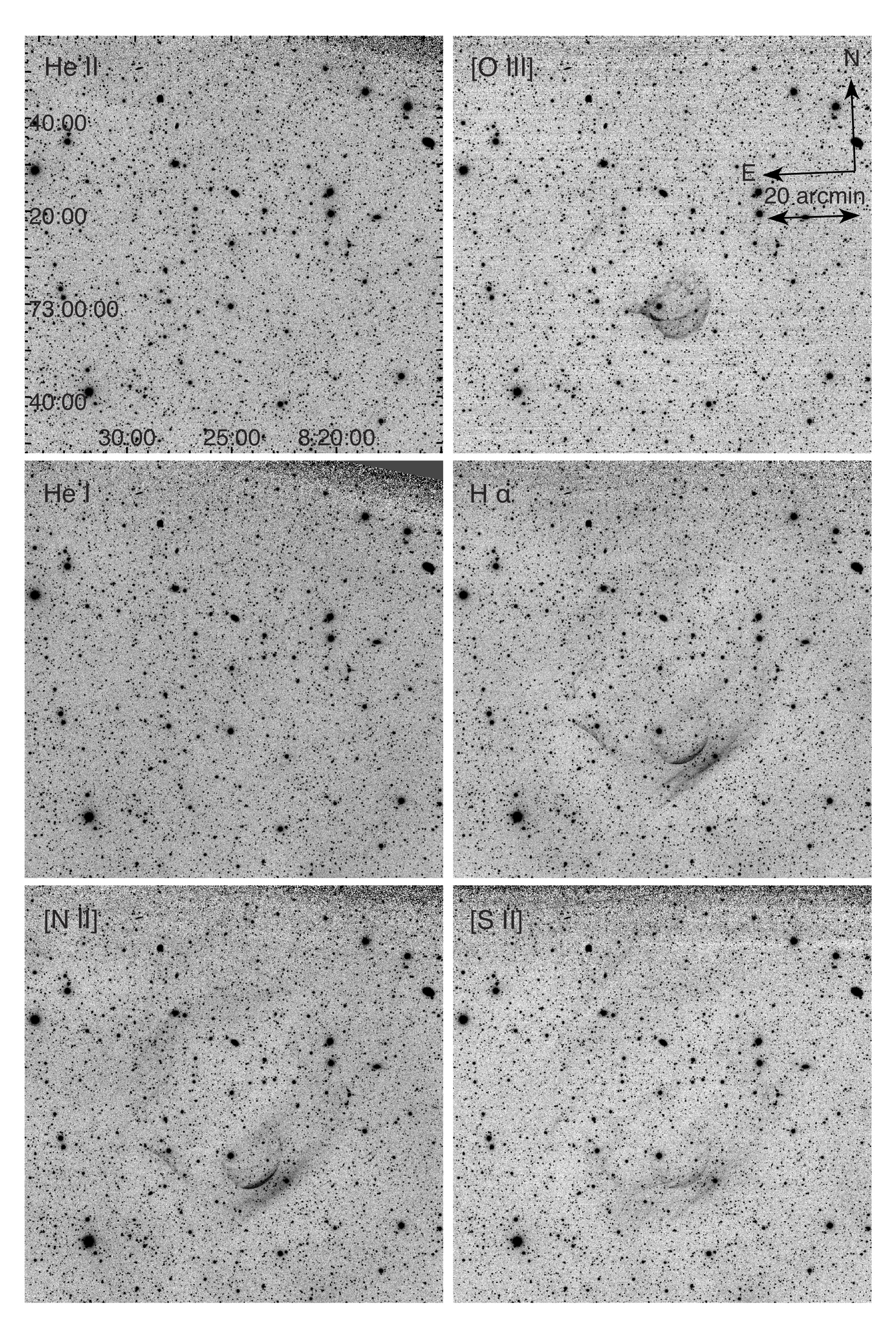}
    \caption{Clockwise from the top right, Condor images of the area surrounding Z Cam in $[\ion{O}{III}]$, H$\,\alpha$, $[\ion{S}{II}]$, $[\ion{N}{II}]$, $\ion{He}{I}$, and $\ion{He}{II}$. The total exposure time in each of the six narrow-band filters was 63,600 sec. Each image was taken with one of the six 180 mm apochromatic refracting telescopes that constitute the Condor Array \citep{Lanzetta2023}. Images have been smoothed via a Gaussian kernel ($\sigma = 2.5$ pixels) and are displayed with linear scaling.}
    \label{fig:six_panel}
\end{figure*}

\begin{figure*}
    \includegraphics[width=\textwidth]{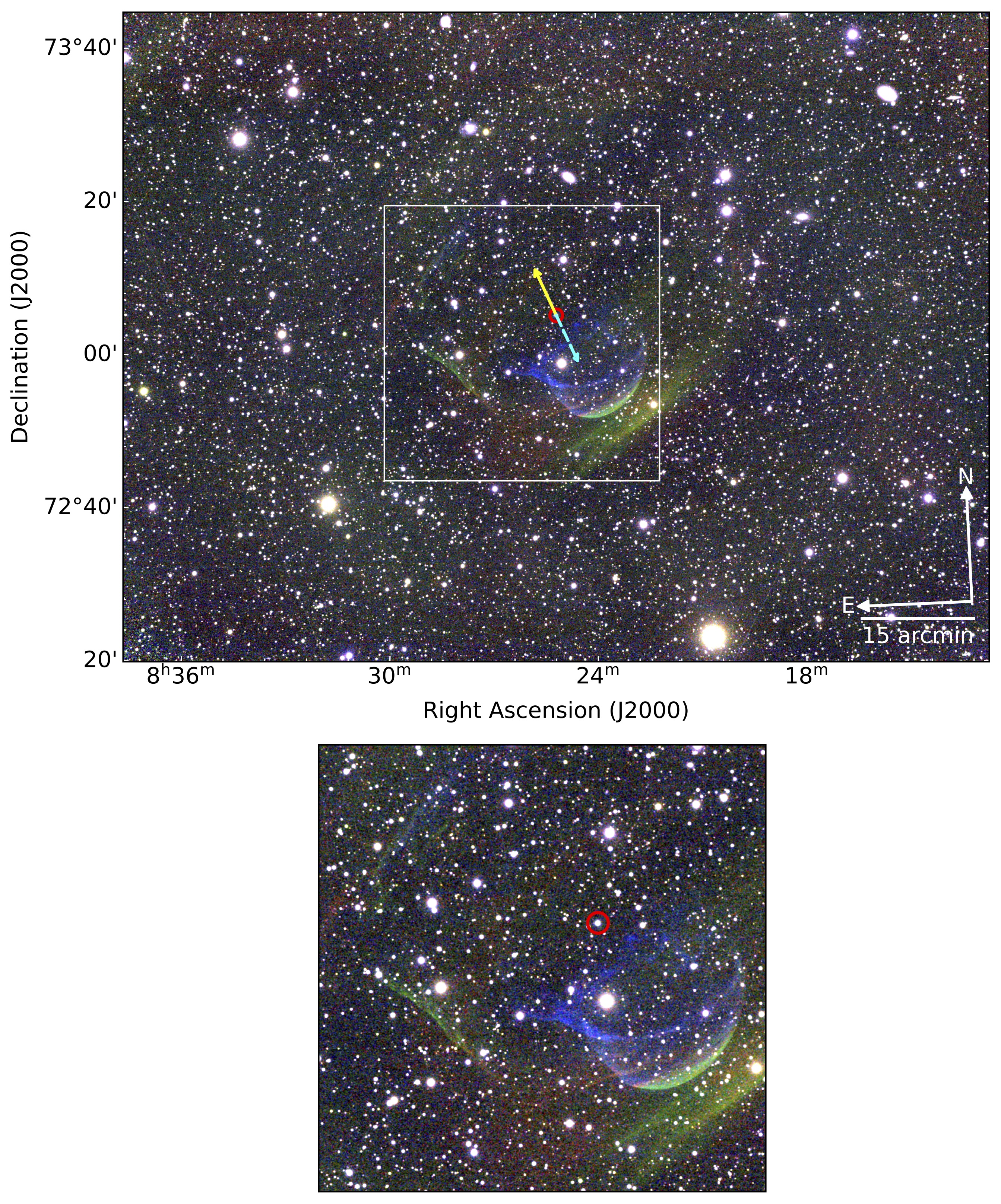}
    \caption{Top: Area surrounding Z Cam in Condor $[\ion{S}{II}]$, H$\,\alpha+[\ion{N}{II}]$, and $[\ion{O}{III}]$ in red, green, and blue, respectively. Z Cam is circled in red. The images have been smoothed via a Gaussian kernel ($\sigma = 2.5$ pixels) and are displayed with asinh scaling. The current direction of motion (from {\it Gaia}, \citealt{Bailer-Jones2021}) of Z Cam is shown with a dashed cyan arrow, and the solid yellow arrow points to the location of Z Cam 20,000 years ago. Bottom: The area of the region containing the SW arc, SE linear feature, and NE linear feature (denoted by the white box) is shown enlarged.}
    \label{fig:color}
\end{figure*}

The six narrow-band images of Z Cam are shown as a mosaic in Fig.~\ref{fig:six_panel}. The $[\ion{S}{II}]$, co-added H$\,\alpha$ and $[\ion{N}{II}]$, and $[\ion{O}{III}]$ images were colorized red, green, and blue, respectively, summed as a single image, and displayed in Fig.~\ref{fig:color}. Despite the fact that the light-collecting area of Condor is $\sim$ 80 times smaller than that of the Mayall 4-m telescope, Condor clearly images fainter features (see Fig.~\ref{fig:color}) than the 4-m telescope (see figs. 1 and 3 of \citealt{Shara2012}) in an exposure time just 1.4 times longer, through filters of comparable transmission efficiency and FWHM. Both the KPNO and Condor images are sky noise limited, so read noise cannot account for the greater Condor sensitivity. It is the combination of Condor's exceptionally clean point spread function and careful image flattening and sky subtraction in the Condor data reduction pipeline that are responsible for the increased sensitivity \citep{Lanzetta2023}. 

In addition to the bright arc to the SW of Z Cam, linear features to the NE, SE, and SW of Z Cam seen in \citet{Shara2012} are clearly visible in the H$\,\alpha$ and $[\ion{N}{II}]$ Condor images (see Fig.~\ref{fig:six_panel}). Notably absent in the Mayall 4-m telescope images is any nebulosity North or NW of Z Cam. A faint arc, brightest in $[\ion{N}{II}]$ and H$\,\alpha$ (green in Fig.~\ref{fig:color}), is now clearly seen in the Condor images to the NE and North of Z Cam. The surface brightness of this arc is $\sim 26.2$ mag arcsec$^{-2}$. This is at least 1 mag arcsec$^{-2}$ more sensitive than the recent search of the environs of 47 CVs \citep{Sahman2022}, which did not locate any new nova shells.

Together with the previously detected nebulosities, this faint arc forms a circular shell $\sim 40$ arcmin in diameter, though notably {\it not} centered on Z Cam itself (circled in red in Fig.~\ref{fig:color}). Radial velocities of the gas throughout the circular shell will be needed to establish with certainty that the northern arc is related to the brighter Z Cam nebulosities, and whether the entire shell is a single entity. 

\begin{figure*}
    \includegraphics[width=\textwidth]{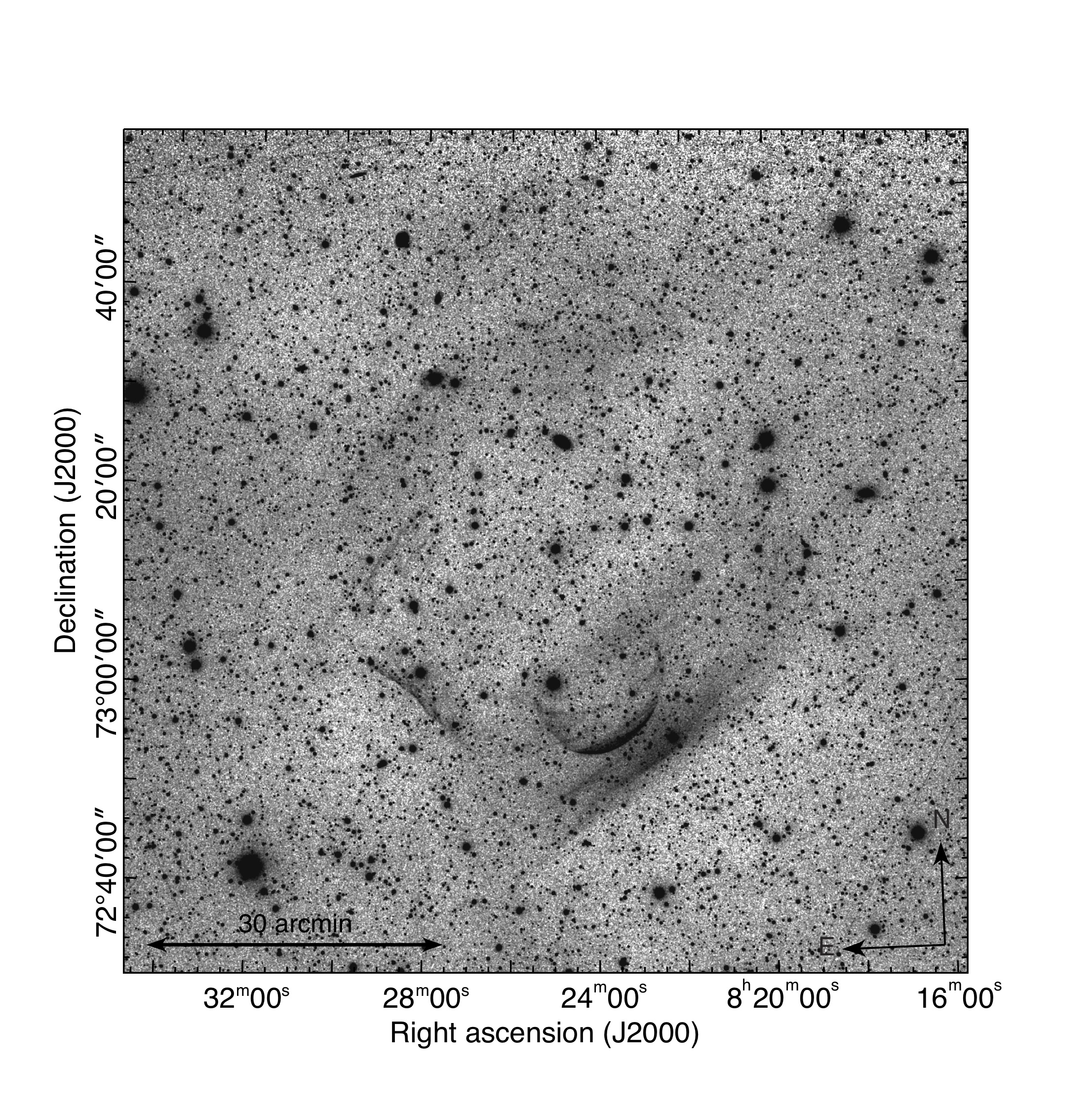}
    \caption{Area surrounding Z Cam in co-added Condor $[\ion{N}{II}]$ and H$\,\alpha$. The image has been smoothed via a Gaussian kernel ($\sigma = 2.5$ pixels) and is displayed with linear scaling. Note the faint nebulosity several arcmin north and east of the NE arc.}
    \label{fig:Halpha_NII}
\end{figure*}

An even fainter arc is seen in the H$\,\alpha$ and $[\ion{N}{II}]$ images a few arcmin north and east of the newly detected NE arc (see Fig.~\ref{fig:Halpha_NII}). If the mass of the WD in Z Cam is $\sim$ 1 M$_\odot$ \citep{Shafter1983}, and its average accretion rate is $\sim 10^{-9} M_\odot$ yr$^{-1}$ \citep{Hartley2005}, then we expect nova eruptions separated by $\sim$16-46 kyr \citep{Yaron2005}. The observed 21.1 mas yr$^{-1}$ proper motion of Z Cam \citep{Bailer-Jones2021} demonstrates that the time between successive nova eruptions is significantly less than the time required ($\sim 100$ kyr) for Z Cam to cross its own shell. It is thus tempting to ascribe the very faint outer arc to a previous eruption of Z Cam. 

\begin{figure*}
    \includegraphics[width=0.8\textwidth]{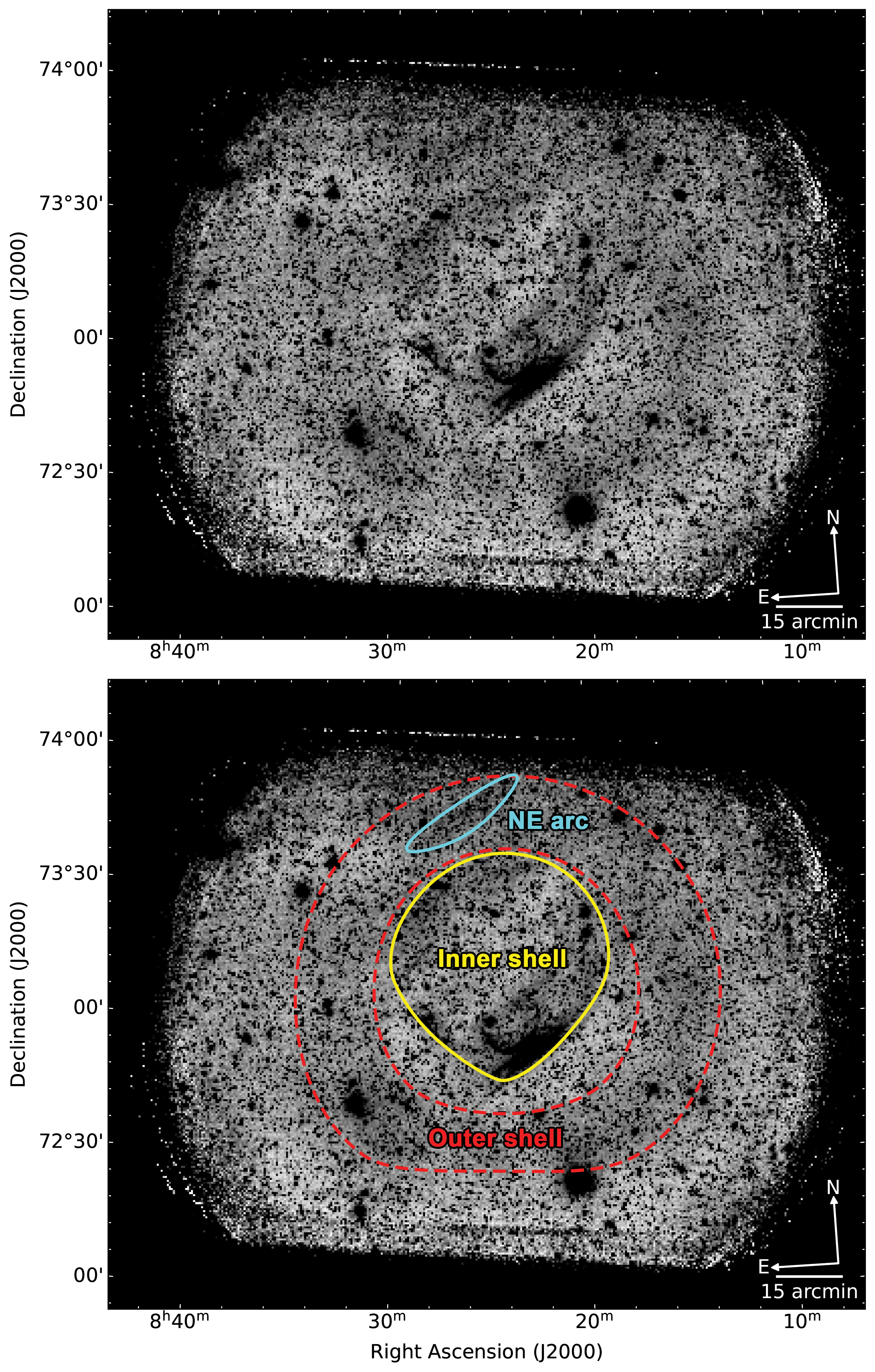}
    \caption[width=\textwidth]{The entire Condor field in co-added $[\ion{N}{II}]$ and H$\,\alpha$ (top) and the same with annotations of notable features overlayed (bottom). The image has been downsampled by a factor of 32 with averaged pixel blocks and is displayed inverted with linear scaling. The now-clear ``inner'' shell encircling Z Cam (``inner shell'', solid yellow outline) is surrounded by an outer, roughly circular nebulosity (though flattened on the southern rim) which is approximately 1.5 degrees in diameter (``outer shell'', between dashed red rings). This larger feature may be the remnant of the nova eruption previous to the one that produced the now fully-seen, inner shell. The nebulosity to the north and east of the inner shell is also visible (``NE arc'', solid light blue outline).}
    \label{fig:Halpha_NII_blocked}
\end{figure*}

In Fig.~\ref{fig:Halpha_NII_blocked} we test this hypothesis by showing the entire Condor field-of-view ($2.5 \times 1.5$ degrees) of the co-added H$\,\alpha+[\ion{N}{II}]$ images, after $10 \times 10$ pixel block averaging. A nearly circular, shell-like structure is seen to encircle the now fully-visible inner shell. No trace of this outer shell is seen in the (equally deep) $\ion{He}{II}$, $\ion{He}{I}$, $[\ion{O}{III}]$ or $[\ion{S}{II}]$ images, or their sum, after $10 \times 10$ pixel binning. This argues strongly against Galactic cirrus as the source of these very low surface brightness features. 

\begin{figure*}
    \includegraphics[width=\textwidth]{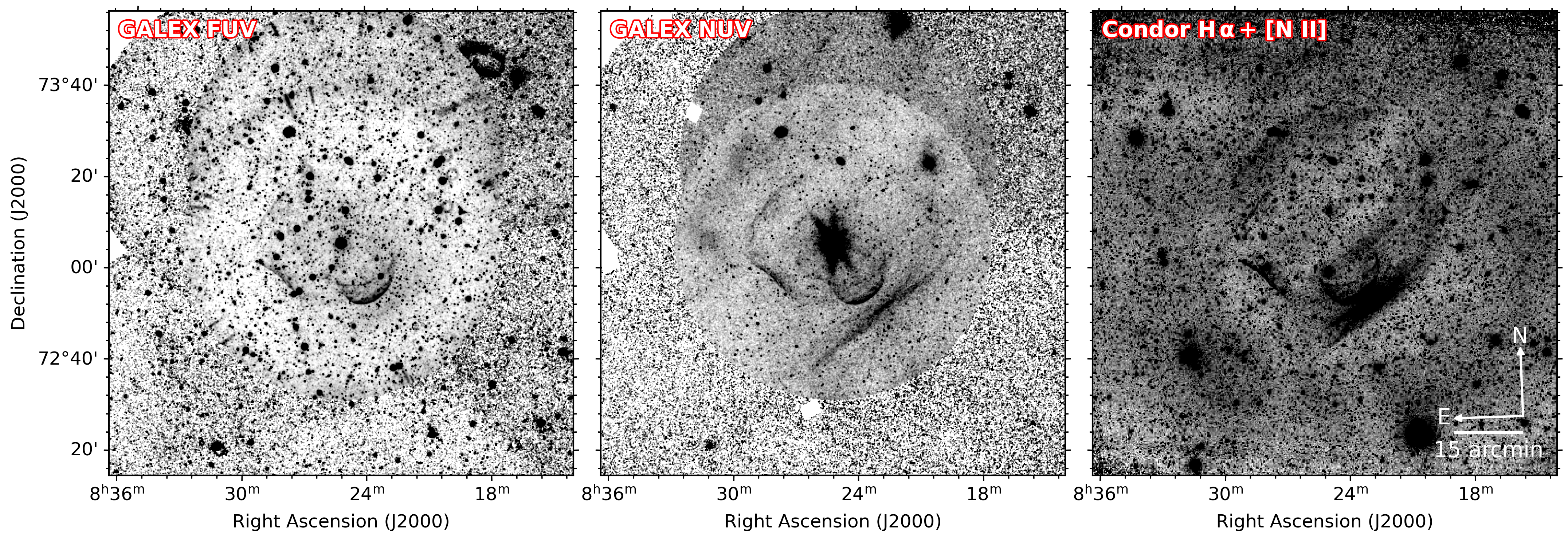}
    \caption[width=\textwidth]{Comparison of \textit{GALEX} FUV, \textit{GALEX} NUV, and Condor H$\,\alpha+[\ion{N}{II}]$ images in the region surrounding Z Cam. The Condor image has been downsampled by a factor of 16 with averaged pixel blocks and is displayed inverted with linear scaling.}
    \label{fig:GALEX_comparison}
\end{figure*}

In Fig.~\ref{fig:GALEX_comparison} we compare the co-added H$\,\alpha+[\ion{N}{II}]$ image with the Galex FUV and NUV archival images. The NE arc is clearly seen in the NUV image, but not the FUV image. The outer shell is not seen in either of the NUV or FUV images.

\begin{figure*}
    \includegraphics[width=\textwidth]{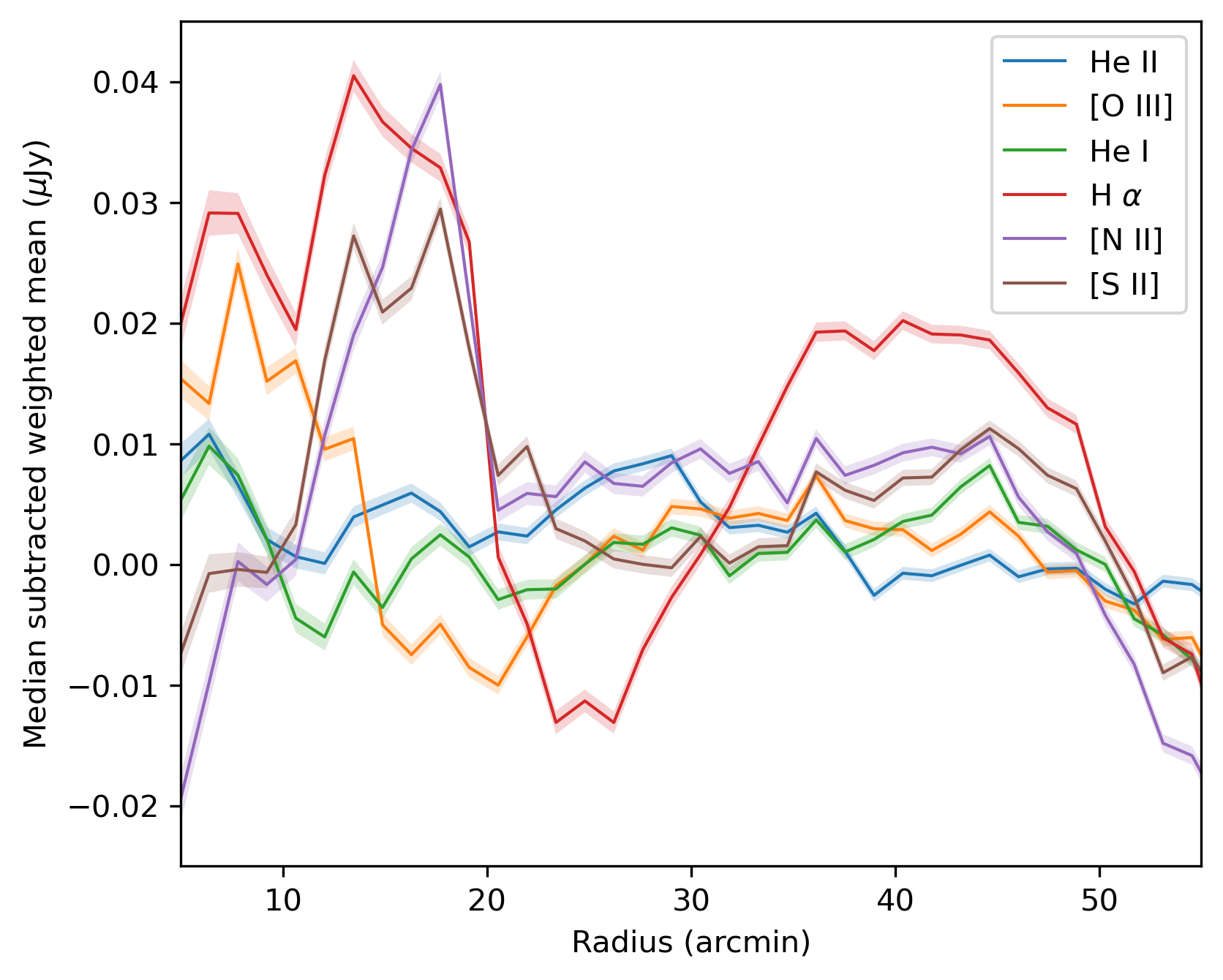}
    \caption{Azimuthally-averaged radial flux profiles for each band, normalized by their median values. The profiles are given by the weighted mean flux per 100 pixel-thick radial bin taken from the approximate center of the outer arc. Standard error on the mean is shown by the shaded regions, which are only slightly thicker than the lines themselves.}
    \label{fig:radial_profiles}
\end{figure*}

In Fig.~\ref{fig:radial_profiles} we show azimuthally-averaged radial flux profiles for each of our six narrow-band images, normalized by their median values. The profiles are given by the weighted mean flux per 100 pixel-thick radial bin taken from the approximate center of the outer, nearly spherical shell. The standard error on the mean is shown by the shaded regions, which are only slightly thicker than the lines themselves. The inner shell, with radius $\sim 15$ arcmin is clearly seen in H$\,\alpha$, $[\ion{N}{II}]$ and $[\ion{S}{II}]$. The outer shell, with radius $\sim$40 arcmin, is strongest in H$\,\alpha$.

Have we detected the first instance of {\it concentric} nova shells due to successive nova eruptions? Nova theory predicts that such nested shells {\it must} exist, and be easiest to detect surrounding massive white dwarfs which erupt on (relatively short, for novae) timescales of order just 1-10 kyr. Though many other such multiple shells should exist, no other shell has ever been imaged to the depth of our images of Z Cam, explaining why it is now the first candidate multiple-shell nova. Deeper imaging and/or spectroscopic characterization of the outer shell will be needed to unambiguously determine its nature.

\section{Measuring the shell expansion}

The brightest and sharpest-edged feature in the complex of nebulosity surrounding Z Cam is the arc $\sim$ 15 arcmin SW of the star \citep{Shara2012}. The 3-year imaging baseline of KPNO images only allowed us to place an upper limit on the expansion of Z Cam's ejecta. The SE quasi-linear feature is too faint in the KPNO and Condor images to yield any expansion estimate.

To quantify the expansion of the SW arc, we employed two methods of determining the spatial shift of nebulosity between the two epochs (2007 and 2021): sum of squared residuals (SSR) minimization and cross-correlation peak measurement. Their means yield our estimate of the shell's velocity, time since eruption, and uncertainties in both these quantities.

\subsection{Image alignment, reprojection, and calibration}

We retrieved a WCS solution for the 2007 Kitt Peak H$\,\alpha+[\ion{N}{II}]$ image from \textsc{astrometry.net} \citep{Lang2010}. Using the adaptive resampling algorithm of \textsc{reproject} \citep{Robitaille2020}, we projected the Kitt Peak image onto the pixel grid of the sum of the Condor H$\,\alpha+[\ion{N}{II}]$ images, matching the latter's 0.85 arcsec/pixel resolution.

To ensure an accurate alignment between the images, we matched sources in the Condor and reprojected Kitt Peak images. With the matched source lists, we used \textsc{astroalign} \citep{Beroiz2020} to determine the affine transformation that matched the Kitt Peak centroids to the Condor centroids, which was then applied to the Kitt Peak image. As noted above, this astrometric matching is accurate to better than 0.1 arcsec across the entire Condor image.

The SW arc (see Fig.~\ref{fig:leading_edge_arc}) was analyzed using a cropped segment of the aligned field. Within this subfield, we subtracted background values for both images and, for SSR minimization, we additionally multiplied the Condor image by a scaling factor determined by the flux of the arc's nebulosity.

\subsection{Sum of squared residuals minimization}

\begin{figure*}
        \includegraphics[width=\textwidth]{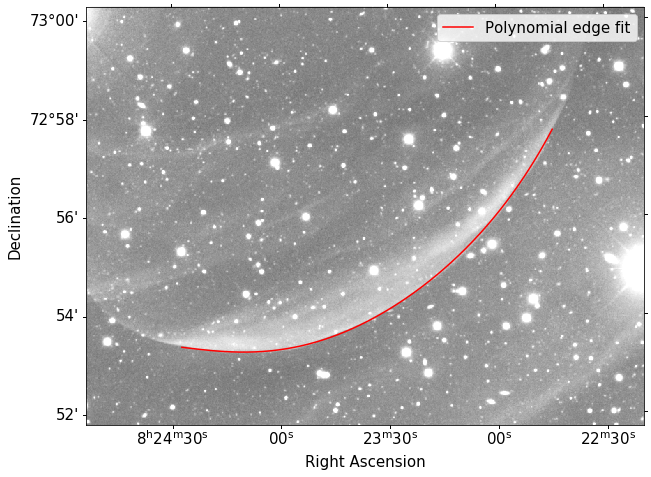}
    \caption{H$\,\alpha+[\ion{N}{II}]$ Kitt Peak image in the region of the SW arc. The leading edge determined by 10th-degree polynomial fitting is shown in red.}
    \label{fig:leading_edge_arc}
\end{figure*}

It is helpful to refer to Fig.~\ref{fig:leading_edge_arc} in the following. To determine the position of the SW arc, we produced vertical flux profiles spanning $\pm15$ pixels of the brightest section of the leading edge in both the Kitt Peak and Condor images. The leading edge was defined via the following procedure. In the Kitt Peak image, we drew by eye a curve of points approximately 10 pixels below the arc. At each point along this curve, we measured the flux moving upwards and recorded the position of first pixel to pass above a threshold (chosen to be approximately half the typical peak flux of the leading edge). This gave a set of points that followed the arc with vertical scatter of $\sim 2$ pixels. We fitted these points with a 10th-degree polynomial and took the result to define the leading edge of the arc. We limited the horizontal range of the edge (i.e. in the Right Ascension direction) to the brightest continuous region for which the process just described gave good polynomial fits. This gave a SW arc edge of length 533 pixels, or 7.64 arcmin (see Fig.~\ref{fig:leading_edge_arc}).

\begin{figure*}
        \includegraphics[width=\textwidth]{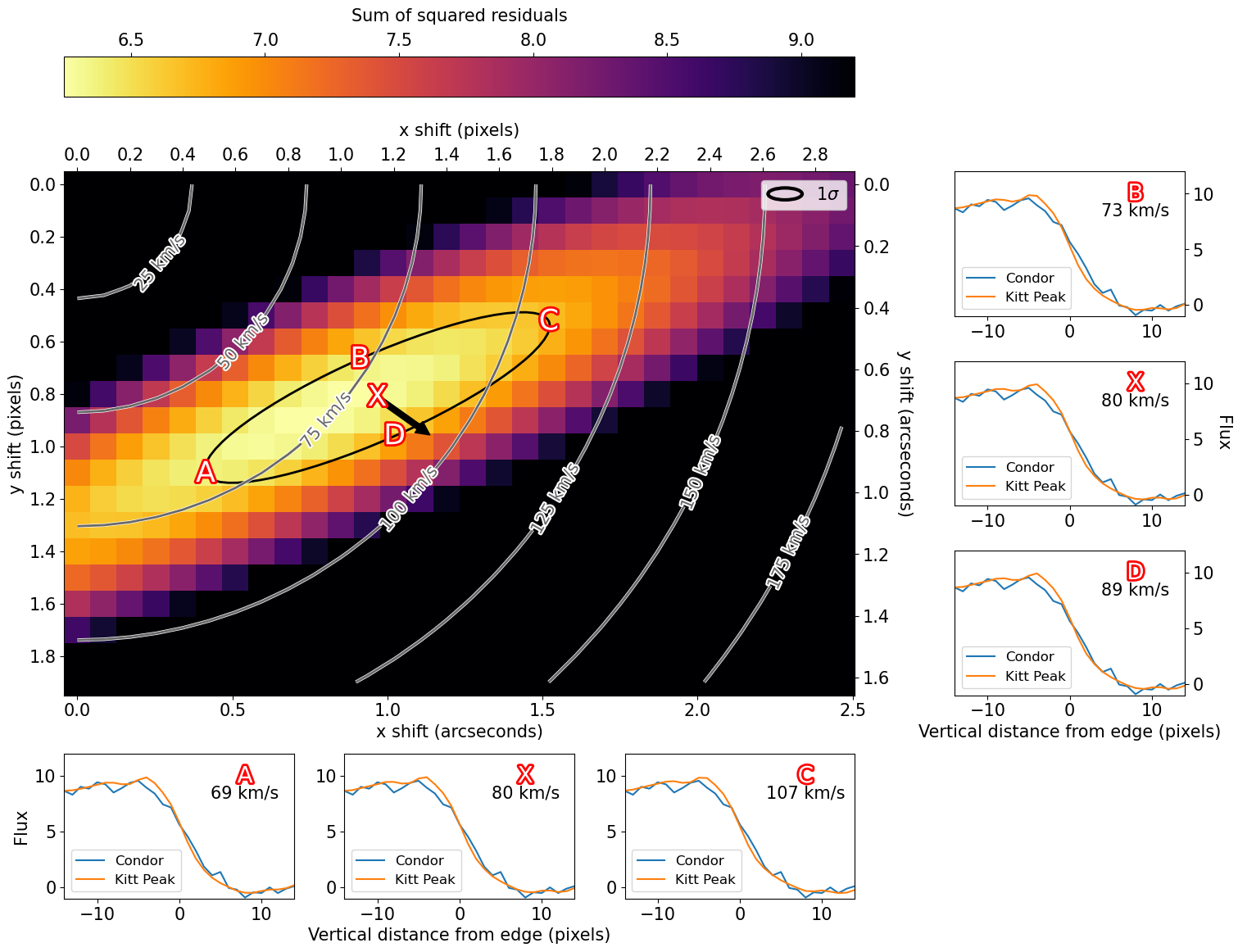}
    \caption{Sum of squared residuals between the Condor and Kitt Peak vertical flux profiles for different pixel shifts of the SW arc, with color data corresponding to the fully sampled leading edge. The point X corresponds to the mean minimum point from 10,000 bootstrapped samples of the arc. The direction of the velocity vector is shown by the black arrow. Vertical profiles for the mean point as well as the vertices of the corresponding error ellipse (A, B, C, and D) are shown below and to the right.}
    \label{fig:ssr_arc}
\end{figure*}

For each pixel along the above-defined leading edge of the arc, we took vertical cuts (i.e. in the Declination direction) spanning $\pm$15 pixels above and below that point. Measuring the flux values at these points for a given image produced 533 vertical profiles each 30 pixels in length, which were then averaged to retrieve a single ``net'' vertical flux profile for the entire arc (see subfigures on the bottom and right of Fig.~\ref{fig:ssr_arc}).



The SSR, calculated as the sum of the the squared differences between the two images' net vertical flux profiles, quantifies the alignment of the feature between the epochs. By shifting the Kitt Peak image in 0.1 pixel increments up to maximum southward and westward shifts of 2.0 pixels and 3.0 pixels and remeasuring the net vertical flux profiles for each shift, we determined the shift for which the SSR is minimized. This coincides with the shift that best aligns the arc in the Kitt Peak image to the arc in the Condor image. This shift gives the net motion of the arc over the nearly 15 year period between the two images and thus the expansion velocity. Image shifting was computed using \textsc{ndimage}'s \citep{Gomers2022} shift function with fifth order spline interpolation. The SSR for the selected shifts is coded by color in the central subfigure of Fig.~\ref{fig:ssr_arc}.

We determined the velocity of the arc by applying this method with bootstrapping \citep{Simpson1986}, randomly selecting 533 points with replacement along the leading edge of length 533 pixels, recording the shift that minimized the SSR, and repeating this process 10,000 times. For some bootstrapped samples, the shift that produced the minimum SSR was found to lie outside of the tested shifts, giving a minimum point on the edge of the search range. We discarded these tests. The mean and standard deviation of the resulting distribution of shifts were taken to represent the most likely shift and a conservative estimate of the uncertainty for this method.


\begin{figure*}
        \includegraphics[width=\textwidth]{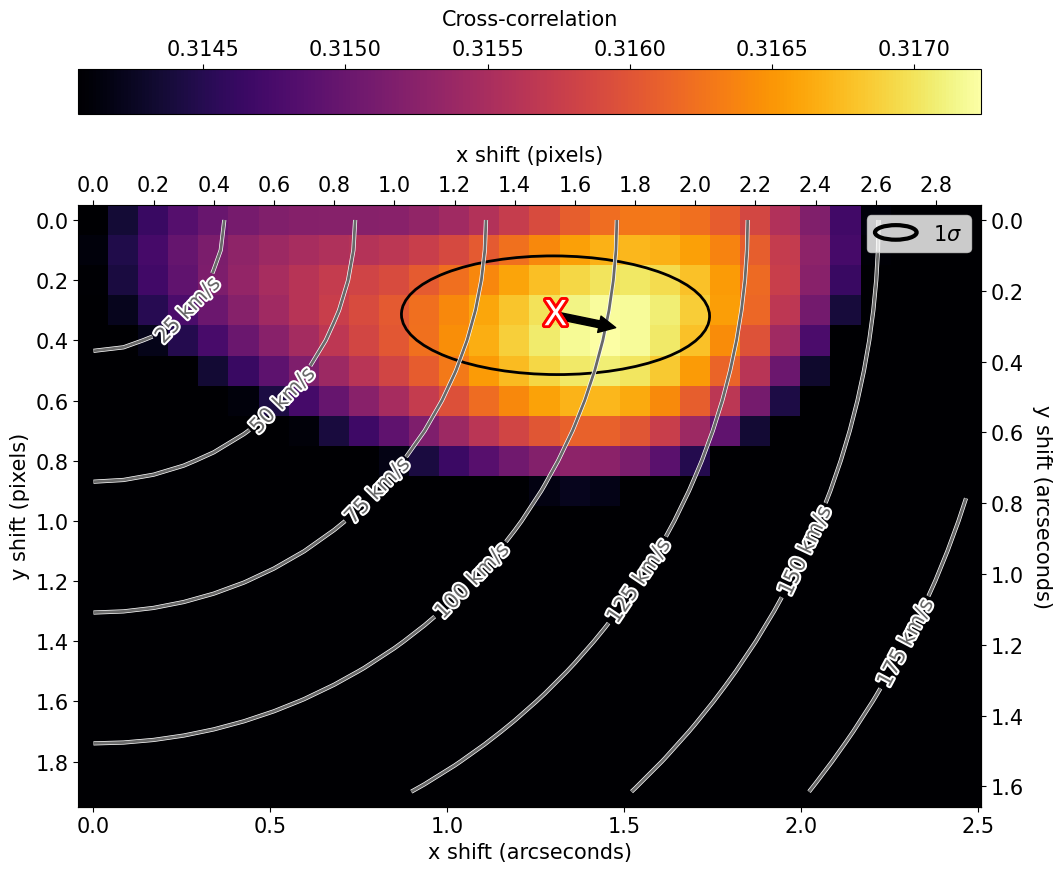}
    \caption{Cross-correlation between the masked Kitt Peak and Condor images for the SW arc. The color boxes correspond to the cross-correlation between the fully-sampled masked images upscaled by a factor of 10. The point X corresponds to the mean maximum point from 10,000 bootstrapped samples at an upscaling factor of 5, with the direction of the velocity vector shown by the black arrow. The black error ellipse is derived from the covariance of the bootstrapped maxima.}
    \label{fig:cc_arc}
\end{figure*}


\subsection{Cross-correlation peak measurement}

Displacement between two images that share a feature can additionally be quantified via cross-correlation. The product of the Fourier transform of an image with the complex conjugate of the Fourier transform of the displaced image yields the Fourier transform of the cross-correlation. Thus the inverse Fourier transform of this product gives the cross-correlation plane, across which coordinates correspond to shifts of one image relative to the other. The position of peak cross-correlation amplitude is representative of the translation that best aligns the features in the two images. Thus we take this point to be the maximum-likelihood shift of the arc between the two epochs. To reduce contaminating the cross-correlation function with sources other than the relevant nebulosity, we masked out regions farther than approximately 10 pixels away from the arc as well as any nearby or overlapping extraneous sources. We calculated the cross-correlation plane using the \textsc{scikit-image} \citep{Vanderwalt2014} masked normalized cross-correlation function \citep{Padfield2012}. To measure cross-correlation on a subpixel scale, we rescaled the images before cross-correlation by a factor of 5. The cross-correlation plane calculated from the images upscaled by a factor of 10 to a 0.1 pixel resolution is coded by color in Fig.~\ref{fig:cc_arc}.

We employed bootstrapping to determine the distribution of likely cross-correlation maxima. This method is advantageous for retrieving measurement statistics on the cross-correlation of distributions with unknown errors without added assumptions \citep{Sciacchitano2019}. This approach has seen astrophysical use in, for example, multiwavelength blazar light curve studies (e.g. \citealt{MaxMoerbeck2014}). As the flux uncertainties of the Kitt Peak image are not known, we limit our implementation of the method (described in \citealt{Peterson1998}) to the ``random subset selection'' procedure. For each of 10,000 iterations, we randomly selected with replacement a number of pixels equal to the total size of the image and removed unselected pixels from the cross-correlation calculation by masking them. Unlike SSR minimisation, in which multiply-selected vertical profiles can proportionally count towards the resulting minimum, multiply-selected pixels in images cannot easily be weighted in their contribution to the cross-correlation. Therefore we instead counted these pixels only once. As this reduces the size of each bootstrapped sample by a factor of approximately $1/e$, we expect the uncertainties to be somewhat larger than a hypothetical fully-sampled bootstrapping method \citep{Peterson1998}. As with the SSR minimisation method, the mean and standard deviation of the distribution of cross-correlation maxima were interpreted as the shift and a conservative estimate of its uncertainty.

\subsection{Determining the motion of the SW arc}

Of the 10,000 bootstrapped samples taken for the SSR minimisation for the SW arc, 7,653 had valid minima within the range of tested shifts. The covariance of the resulting points produced the error ellipse shown in Fig.~\ref{fig:ssr_arc}. Z Cam is the {\it Gaia} DR3 source 1123169888190445568, located at a distance of 214\,pc \citep{Bailer-Jones2021}. These data suggest the SW arc is expanding to the west and south at $65\pm38$\,km\,s$^{-1}$ and $47\pm19$\,km\,s$^{-1}$, respectively. This corresponds to a net motion of $80\pm42$\,km\,s$^{-1}$ at $36\degr$ south of west.

The mean cross-correlation peak of the SW arc appears at a west and south translation corresponding to $88\pm29$\,km\,s$^{-1}$ and $18\pm11$\,km\,s$^{-1}$, respectively, or a net motion of $90\pm32$\,km\,s$^{-1}$ at $12\degr$ south of west (see Fig.~\ref{fig:cc_arc}).

The combined result of these two methods, a shift and formal uncertainty as the means of the two resulting shifts and uncertainties, gives the expansion of the SW arc to be a net motion of $83\pm37$\,km\,s$^{-1}$ at $23\degr$ south of west.

Taking the time since the eruption of Z Cam given by ejecta in the snowplow phase to be $t=r/4v$ \citep{Oort1951}, where the distance from Z Cam to the SW arc is $r=0.9$\,pc, we find $t=2772$ years via SSR minimisation, and $t=2473$ years via the cross-correlation peak. The modest difference between these two ages (just 299 years, or 11\% of the age we adopt below) is encouraging, and suggests that the formal errors associated with each method above may be significant overestimates. We nonetheless adopt the mean value and mean uncertainty for our adopted age of the Z Cam shell: $2672^{-817}_{+2102}$ yr.

\section{Conclusions}
Deep narrow-band images of the ejecta of Z Cam, taken with the Condor Array Telescope, have revealed the nearly complete, circular shell, $\sim$ 30 arcmin in diameter, of this old nova/dwarf nova. An $\sim$80 arcmin diameter, fainter concentric shell is also detected. If the faint outer shell is confirmed, Z Cam would be the first instance of a {\it pair} of faint, concentric shells surrounding an old nova, representing the two most recent eruptions of the system's white dwarf.   

We have measured the expansion velocity of the brightest ejecta to be $83 \pm 37$\,km\,s$^{-1}$, and its age to be $2672^{-817}_{+2102}$ yr. The suggestion that the ``new'' star first seen by Chinese Imperial astrologers in 77 BCE \citep{Ho1962} was a nova eruption of Z Cam \citep{Johansson2007} is consistent with our deduced expansion age of Z Cam's ejecta. (We also cannot rule out all the suggestions of \citealt{Warner2015} of more recent transients, but see also \citealt{Hoffmann2019,Hoffmann2020} and \citealt{Neuhauser2021}). The age uncertainty of the last eruption of Z Cam remains large, so the identity of the underlying star of the transient of 77 BCE is far from settled. 

\section*{Acknowledgements}
MMS, KML and JTG gratefully acknowledge the support of NSF award 2108234. This material is based upon work supported by the National Science Foundation under Grants 1910001, 2107954, and 2108234. Based on observations at Kitt Peak National Observatory, NSF's National Optical Astronomy Observatories (NOAO Prop. 2006B-0417; PI:Shara), which is operated by the Association of Universities for Research in Astronomy (AURA) under a cooperative agreement with the National Science Foundation. This work has made use of data from the European Space Agency (ESA) mission {\it Gaia} (\url{https://www.cosmos.esa.int/gaia}), processed by the {\it Gaia} Data Processing and Analysis Consortium (DPAC,
\url{https://www.cosmos.esa.int/web/gaia/dpac/consortium}). Funding for the DPAC has been provided by national institutions, in particular the institutions
participating in the {\it Gaia} Multilateral Agreement.
MMS acknowledges the longtime interest of Göran Johansson, who first suggested that Z Cam was the nova of 77 BCE, and correspondence with S. Hoffmann and R. Neühauser about historical novae.

Facilities: KPNO, Condor

Software: \textsc{astroalign} \citep{Beroiz2020}, \textsc{astrometry.net} \citep{Lang2010}, \textsc{astropy} \citep{astropy:2013, astropy:2018, astropy:2022}, \textsc{daophot} \citep{Stetson1987}, \textsc{ndimage} \citep{Gomers2022}, \textsc{numpy} \citep{harris2020array}, \textsc{reproject} \citep{Robitaille2020}, \textsc{scikit-image} \citep{Vanderwalt2014}

\section*{Data Availability}

All relevant data, including all images and figures,
are available from the corresponding author on reasonable request.

\newpage    



\bibliographystyle{mnras}
\bibliography{zcam} 







\bsp	
\label{lastpage}
\end{document}